\documentclass[12pt]{article}
\pdfoutput=1
\usepackage{amsmath}
\usepackage{amssymb}
\usepackage{graphicx}
\usepackage{subcaption}
\usepackage{url}

\usepackage[sort&compress, numbers, merge]{natbib}

\usepackage[colorlinks=true, linkcolor=blue, citecolor=blue,
urlcolor=black]{hyperref}

\begin{document}

\begin{titlepage}
\begin{center}

\hfill UCB-PTH-15/04\\
\hfill DESY 15-105 \\
\hfill UT-15-21 \\
\vspace{2mm}
\hfill \today

\vspace{3.0cm}
{\large\bf ATLAS $Z$ Excess in Minimal Supersymmetric Standard Model}

\vspace{1.0cm}
{\bf Xiaochuan Lu}$^{(a)}$,
{\bf Satoshi Shirai}$^{(b)}$ and
{\bf Takahiro Terada}$^{(c,\, b)}$

\vspace{1.0cm}
{\it
$^{(a)}${ Berkeley Center for Theoretical Physics, Department of Physics, \\
     and Theoretical Physics Group, Lawrence Berkeley National Laboratory, \\
     University of California, Berkeley, CA 94720, USA}\\
$^{(b)}${Deutsches Elektronen-Synchrotron (DESY), 22607 Hamburg, Germany} \\
$^{(c)}${Department of Physics, University of Tokyo, Tokyo 113-0033,Japan}
}

\vspace{1cm}
\abstract{
Recently the ATLAS collaboration reported a 3$\sigma$ excess in the search for the events containing a dilepton pair from a $Z$ boson and large missing transverse energy. Although the excess is not sufficiently significant yet, it is quite tempting to explain this excess by a well-motivated model beyond the standard model. In this paper we study a possibility of the minimal supersymmetric standard model (MSSM) for this excess.
Especially, we focus on the MSSM spectrum where the sfermions are heavier than the gauginos and Higgsinos.
We show that the excess can be explained by the reasonable MSSM mass spectrum.
}
\end{center}
\end{titlepage}

\setcounter{footnote}{0}

\section{Introduction}
Recently the ATLAS collaboration has reported a 3$\sigma$ excess in the search for the events with a dilepton pair on $Z$ boson mass peak and large missing transverse energy (MET) \cite{Aad:2015wqa}.
The signal events should contain a same-flavor opposite-sign dilepton pair with its invariant mass in the $Z$ boson mass range,  large MET ($E_{\text{T}}^{\text{miss}}>225$ GeV),
and large scalar sum ($H_{\text{T}}>600$ GeV) of the transverse momenta of all signal jets and the two leading leptons.
The observed events in the dielectron and dimuon channels are 16 and 13 respectively, whereas the expected numbers of the standard model (SM) background events are $4.2\pm1.6$ and $6.4\pm 2.2$.  The two channels combine to give 29 observed events compared to $10.6\pm 3.2$ expected from the SM, amounting to a 3$\sigma$ excess.

On the other hand, the CMS collaboration has an analogous search for events with large MET and a dilepton pair on $Z$ \cite{Khachatryan:2015lwa}, where the signal events are classified into MET bins ($E_{\text{T}}^{\text{miss}} = 100$-$200$, 200-300 and $>300$ GeV) and the number of jets $n_{\rm jet}\geq2,3$.
In this CMS counterpart search, no significant deviation from the SM expectation is found.

Although the present ATLAS $Z$ excess is not so statistically significant yet and might even conflict with the aforementioned counterpart search by the CMS collaboration, it is quite tempting to investigate whether a well-motivated model beyond the SM can explain the signal.
The large MET  is a typical signature of the supersymmetric (SUSY) SMs.
Several studies are devoted to the ATLAS $Z$ excess in the context of the SUSY SMs.

In Ref.~\cite{Barenboim:2015afa}, general requirements in the SUSY SMs to explain the excess are studied, and it is concluded that there needs to be a particle lighter than about 1.2 TeV,  with  production cross section the order of colored particles, as well as producing $O(1)$ $Z$ bosons in its decay chain.
In the specific case of the minimal SUSY standard model (MSSM) with a neutralino as the lightest SUSY particle (LSP), it is found that $Z$ bosons are generically not produced enough in the SUSY cascade decay chain.
One can try to compensate this by increasing the production cross section with the light gluinos or squarks, 
but the constraints from the jets and MET searches~\cite{Aad:2014wea} get more severe.
Possible solutions to overcome this difficulty are discussed, including use of a compressed spectrum or a spectrum with the light gravitino LSP, in which the neutralino next-to-LSP (NLSP) decays into the gravitino and $Z$ boson.
Another type of spectrum is studied in Ref.~\cite{Cahill-Rowley:2015cha}, where the first and second generation squarks decay into a Bino followed by the Bino decaying into Higgsinos and $W$, $Z$ boson, or the Higgs boson $h$.

The spectrum with the light gravitino LSP is realized in the so-called general gauge mediation (GGM)~\cite{Meade:2008wd, Buican:2008ws}, where the branching fraction of neutralino NLSP into $Z$ can be close to one.
However, constraints from other SUSY searches such as jets$+$MET~\cite{Aad:2014wea}, stop search~\cite{Aad:2014qaa} or multi-lepton~\cite{Aad:2014iza,Chatrchyan:2014aea} as well as CMS on-$Z$ dilepton~\cite{Khachatryan:2015lwa} are severe and such an explanation is not viable~\cite{Allanach:2015xga}.

On the other hand, compressed spectra are utilized in other attempts to explain the excess~\cite{Ellwanger:2015hva, Kobakhidze:2015dra, Cao:2015ara}.
When the mass difference between the neutralino LSP and the neutralino NLSP is less than the Higgs mass (125 GeV), the two-body decay of the NLSP into the LSP plus a $Z$ boson can be efficient.
An important requirement here is to ensure that the parent particle (gluino or squark) decays mainly into the NLSP so that the NLSP can produce $Z$ in the next step of the decay chain.
In the case of the GGM with a light gravitino, this requirement is satisfied because the gravitino LSP is very weakly coupled.
In the heavy LSP scenarios, the LSP is taken to be a Bino-like neutralino in the MSSM with light sbottom~\cite{Kobakhidze:2015dra} or singlino-like in the next-to-MSSM (NMSSM)~\cite{Ellwanger:2015hva, Cao:2015ara}.
Combining other constraints~\cite{Aad:2013wta, Aad:2014wea, Khachatryan:2015lwa}, the NMSSM scenario can reduce the significance of the excess only in a small region in the parameter space~\cite{Cao:2015ara}.
The light sbottom scenario~\cite{Kobakhidze:2015dra} can also reduce the significance, but sbottom produces bottom quarks when it decays into Higgsino-like neutralinos, and hence is severely constrained by $b$-jets searches~\cite{Aad:2014lra}.
There is also a non-SUSY study~\cite{Vignaroli:2015ama} in the composite Higgs~\cite{Kaplan:1983fs} / Randall-Sundrum~\cite{Randall:1999ee} framework, but this also has bottom-rich signatures.

In this paper, we revisit the possibility of explaining the excess in the MSSM, in particular in the well-motivated split SUSY-like spectrum.
The MSSM is one of the most attractive candidates of models beyond the standard model.
Especially the recent discovery of the Higgs boson $h$ with a mass of around 125 GeV \cite{Aad:2012tfa,*Chatrchyan:2012ufa} seems to suggest the framework of the split SUSY~\cite{Wells:2003tf,*Wells:2004di,
ArkaniHamed:2004fb,*Giudice:2004tc,*ArkaniHamed:2004yi,*ArkaniHamed:2005yv}, where the SUSY fermions are around TeV scale and the SUSY scalars are heavier than TeV scale.
This framework can overcome weak points of the weak-scale MSSM, such as SUSY flavor/CP and cosmological problems.
Most importantly, the split SUSY is quite compatible with the observed Higgs mass \cite{Okada:1990vk,*Okada:1990gg,*Ellis:1990nz,*Haber:1990aw,*Ellis:1991zd,Giudice:2011cg}.
In the light of the Higgs discovery, this framework is intensively studied~\cite{Hall:2011jd,*Hall:2012zp,*Nomura:2014asa, Ibe:2011aa,
*Ibe:2012hu, Arvanitaki:2012ps, ArkaniHamed:2012gw}.
Now the MSSM with split SUSY-like spectrum is one of the most convincing and  viable models.
Therefore it is very interesting to study this model in light of the ATLAS $Z$ excess.

To explain the ATLAS $Z$ excess, the MSSM mass spectrum will at least satisfy the condition that the SUSY cascade decay chain has large branching fraction to $Z$.
Even if the $Z$-rich decay chains are realized at the LHC,  
the non-leptonic decays of the $Z$ bosons induce signals of 
multi-jets plus MET with zero lepton, which are severely constrained.
Moreover, multiple leptonic decays of the $Z$ bosons may result in multi-lepton events.
Therefore our goal is to find the mass spectra which satisfy the conditions:
\begin{itemize}
\item SUSY cascade decay chain is $Z$-rich. 
\item Less constrained by searches other than the dilepton channel, such as multi-jets+MET.
\end{itemize}

\begin{figure}[t]
  \centering
  \includegraphics[width=0.7\textwidth]{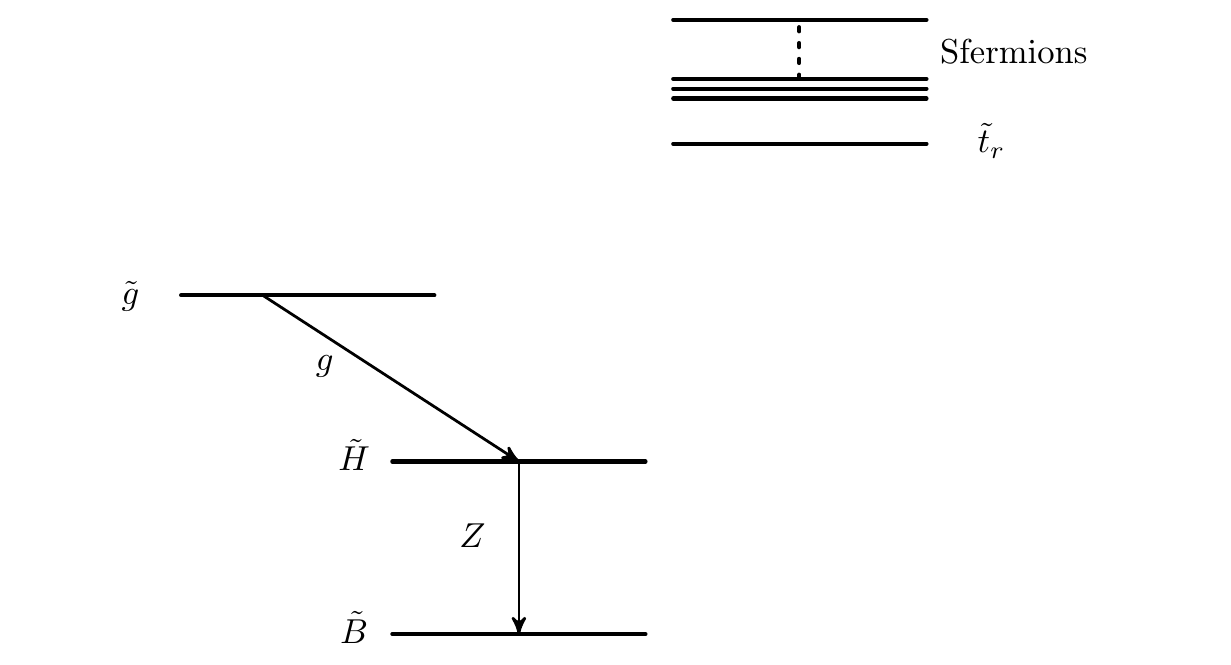}
  \caption{ \label{fig:spectrum} Schematic picture of the present MSSM spectrum.
  }
   \end{figure}
In this paper, we point out that a simple mass spectrum (Fig.~\ref{fig:spectrum}) can well account for the ATLAS $Z$ excess.
If the mass spectrum is compressed enough, $m_{\tilde H} - m_{\tilde B}\sim 100$ GeV and $m_{\tilde H} \gtrsim m_{\tilde g} - 2 m_{t}$,
the above conditions can be simultaneously satisfied.
As we will see in Section~\ref{sec:spectrum}, when the stop is heavy and the spectrum is sufficiently compressed, the gluino radiative decay into a Higgsino-like neutralino and a gluon becomes the dominant gluino decay channel.
The Higgsino-like neutralino can then decay into $Z$ boson with a branching fraction close to 1.
Together, this gives rise to an efficient production of $Z$ boson.
In Section~\ref{sec:signals}, we reduce essential features of the MSSM spectrum to a simplified model, and study LHC signals of the model and its constraints.
Summary and discussions are given in Section~\ref{sec:summary}.

\section{SUSY Spectrum}\label{sec:spectrum}
\subsection{Mass spectrum}

We take the split SUSY-like spectrum, where the gauginos and Higgsinos are light whereas scalar superparticles are heavy, and we consider production of the relatively light gluino, which decays into neutralinos and charginos.
The LSP is a Bino-like neutralino $\tilde{\chi}_1^0$, and there are nearly degenerate two Higgsino-like neutralinos $\tilde{\chi}_{2, 3}^0$ and a Higgsino-like chargino $\tilde{\chi}^{\pm}_1$ as the NLSPs.  For simplicity, we take the Wino heavier than the gluino.

\begin{figure}[htbp]
\centering
   \includegraphics[width=0.5\textwidth]{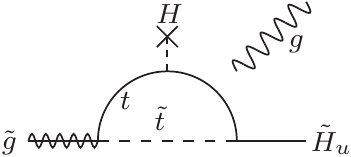}
  \caption{\label{fig:gluino_decay} Diagrams of the gluino decay into a Higgsino and gluon.}
   \end{figure}

Because of the renormalization group effects, the right-handed stop is expected to be typically lighter than the other squarks.
If the right-handed stop is lighter among the squarks, gluino branching fraction shows an interesting feature: the dominant decay channel becomes the top-stop-loop-induced process into a Higgsino-like neutralino and a gluon for suitable mass splitting between gluino and the neutralinos~\cite{Toharia:2005gm, Gambino:2005eh, Sato:2012xf,*Sato:2013bta} (see Fig.~\ref{fig:gluino_decay}).
This reduces the branching fraction of gluino three-body decay modes, and the gluino dominantly decays into a neutral Higgsino with a gluon.

\subsection{Decay of neutralinos and gluino}

We study neutralino and gluino decays in this Subsection to motivate the present split SUSY spectrum and a simplified model studied in the next Section.

\subsubsection*{Decay of neutralinos}
First, we consider the neutralino decay.
A neutralino may decay into a lighter neutralino emitting a $Z$ or $h$ boson, and into a chargino with a $W$ boson if each channel is kinematically allowed.
In the present mass spectrum, the lightest chargino $\tilde{\chi}^{\pm}_1$ is Higgsino-like, so it is approximately degenerate with Higgsino-like neutralinos $\tilde{\chi}_{2, 3}^{0}$.
This means that neutralinos cannot decay into the chargino with a $W$ boson in the present spectrum.
If the mass difference between  $\tilde{\chi}_1^0$ and $\tilde{\chi}_{2, 3}^0$ is greater than $m_{Z}$ and less than $m_h$,  $\text{BF}(\tilde\chi_{2, 3}^0 \to \tilde\chi_1^0 Z)\simeq 1$.
Once the Higgs channel becomes kinematically available, the branching ratio of each Higgsino-like neutralino into $Z$ or $h$ varies substantially in the case of low tan$\beta$ depending on the phases of parameters such as $\mu$-term, but 
 $\sum_{i=2, 3}\text{BF}(\tilde{\chi}_i^0 \to \tilde{\chi}_1^0 Z) /2 \simeq \sum_{i=2, 3}\text{BF}(\tilde{\chi}_i^0\to \tilde{\chi}_1^0 h) /2  \simeq 0.5 $ 
  in the limit of large mass difference between $\tilde{\chi}_1^0$ and $\tilde{\chi}_{2, 3}^0$.
This can be understood in the Nambu-Goldstone picture: the longitudinal component of the $Z$ boson is the complex partner of the Higgs, whereas the transverse components are unimportant in the limit.
Taking the average is justified because gluino decays into the up-type Higgsino in our spectrum (see below), and it is approximately equally contained in the two mass eigenstates of Higgsino-like neutralinos.
Therefore in order to produce the $Z$ bosons in the gluino decay chain efficiently,
$m_Z <m_{\tilde{\chi}^0_{2, 3}}-m_{\tilde{\chi}^0_1} \lesssim m_h$ is required.
In the following analysis, we assume $m_{\tilde{\chi}^0_{2, 3}}-m_{\tilde{\chi}^0_1}\simeq 100$ GeV.

\subsubsection*{Decay of gluino}
Next, let us move on to the gluino decay.
As studied in Refs.~\cite{Toharia:2005gm, Gambino:2005eh, Sato:2012xf,*Sato:2013bta}, the partial decay rate of gluino into a gluon and an (up-type) Higgsino is relatively enhanced by a factor $\left( \log \left( m_{\tilde{t}}/m_{t} \right) \right)^2$ compared to other channels: (i) gluon and Bino, (ii) neutralino, quark and antiquark, and (iii) chargino, quark and antiquark.
Therefore, the gluino efficiently produces Higgsino-like neutralinos $\tilde{\chi}^0_2$ and $\tilde{\chi}^0_3$ in the case of the light gluino and heavy (but relatively lighter among the squarks) stop.
This radiative decay of the gluino has some advantages to explain the ATLAS $Z$ excess.
The gluino decay into a Higgsino-like neutralino, followed by the decay of the neutralino into a $Z$ boson, efficiently produces the $Z$ boson.
Thanks to the log enhancement, this decay mode dominates over the other channels for wide parameter space and the expectation value of the number of $Z$ bosons per gluino decay is enhanced.
In this case, the constraints from other SUSY searches, such as multi-jets and leptons signals, can be relaxed.
Another advantage of this radiative decay is that we can suppress the branching fractions into heavy-flavor jets, which are severely constrained by LHC searches even for the compressed mass spectrum.

In the estimation of gluino branching fractions, we need resummation of the $ \log \left( m_{\tilde{t}}/m_{t} \right) $ factor, otherwise, the two body decay rate is overestimated \cite{Gambino:2005eh}.
For this resummation, we first evaluate the Wilson coefficients of the dimension five dipole operator $O^{\tilde{B}}_{7} \equiv ({{\tilde B}} \sigma^{\mu\nu}  {\tilde g}) G_{\mu\nu}$, 
 dimension six dipole operator $O^{\tilde{H}_u}_{5} \equiv ({ {\tilde H_u}} \sigma^{\mu\nu}  {\tilde g}) H G_{\mu\nu}$, and several four-Fermi operators which include a gluino spinor e.g., $O^{\tilde{H}_u}_{2,ij}\equiv (\tilde H_u  \sigma^{\mu\nu}  {\tilde g} ) (Q_{L,i} \sigma_{\mu\nu}u^c_{R,j} )  $, at the sfermion mass scale.
The operators relevant for the radiative decay of the gluino are 
$O^{\tilde{B}}_{7}$,  $O^{\tilde{H}_u}_{5}$, and $O^{\tilde{H}_u}_{2,33}$.
At the sfermion mass scale, these Wilson coefficients are given by
\begin{align}
C^{\tilde{B}}_{7}& = \frac{g_s^2 g'}{384\pi^2} (m_{\tilde g} - m_{\tilde B})\sum_i
\left( \frac{1}{m^2_{\tilde{q}_{Li}}} 
-  \frac{2}{m^2_{\tilde{u}_{Ri}}}  +  \frac{1}{m^2_{\tilde{d}_{Ri}}} 
    \right),\\
C^{\tilde{H}_u}_{2,33}& = \frac{g_s y_t}{4\sqrt{2}\sin\beta}
 \left(\frac{1}{m^2_{\tilde{q}_{L3}}} + \frac{1}{m^2_{\tilde{t}_{R}}} \right),\\
C^{\tilde{H}_u}_{5}& =  \frac{g_s^2 y_t^2}{32\sqrt{2}\pi^2\sin\beta}
 \left(\frac{1}{m^2_{\tilde{q}_{L3}}} + \frac{1}{m^2_{\tilde{t}_{R}}} \right).
\end{align}
Here we neglect sfermion flavor violation and the Yukawa couplings other than $y_t$.
 By using the renormalization group equations (RGEs) of the Wilson coefficients, we evolve down to the gluino mass scale.
The RGEs of interest are written as,
\begin{align}
\frac{d}{d\ln\mu}C^{\tilde{B}}_{7} &= \frac{-14g_s^2}{16\pi^2}C^{\tilde{B}}_{7},\\
\frac{d}{d\ln\mu
}\begin{pmatrix}
C^{\tilde{H}_u}_{2,33}\\
C^{\tilde{H}_u}_{5}
\end{pmatrix}
&= \frac{1}{16\pi^2}
\begin{pmatrix}
-\frac{37}{3}g_s^2 + \frac{3}{2}y_t^2 & 2 g_s y_t \\
4g_sy_t & -14g_s^2 + 3 y_t^2
\end{pmatrix}
\begin{pmatrix}
C^{\tilde{H}_u}_{2,33}\\
C^{\tilde{H}_u}_{5}
\end{pmatrix}.
\end{align}
Here we take into account only the top Yukawa and strong couplings.
The Wilson coefficient $C^{\tilde{H}_u}_{5}$ is logarithmically enhanced through the operator mixing with $O^{\tilde{H}_u}_{2,33}$ and $C^{\tilde{B}}_{7}$ is not.
Thus, the radiative decay rate of the gluino into the Higgsino are relatively enhanced with heavier sfermions.
We evaluate the gluino decay rates with these Wilson coefficients at the gluino mass scale.
For details, see Refs.~\cite{Gambino:2005eh, Sato:2012xf,*Sato:2013bta}. 
Following Ref. \cite{Buttazzo:2013uya}, we set the weak scale Yukawa and gauge coupling parameters.

\begin{figure}[h!]
 \centering
 \subcaptionbox{Branching fraction of each decay mode \label{subfig:br1}}{\includegraphics[width=0.45\textwidth]{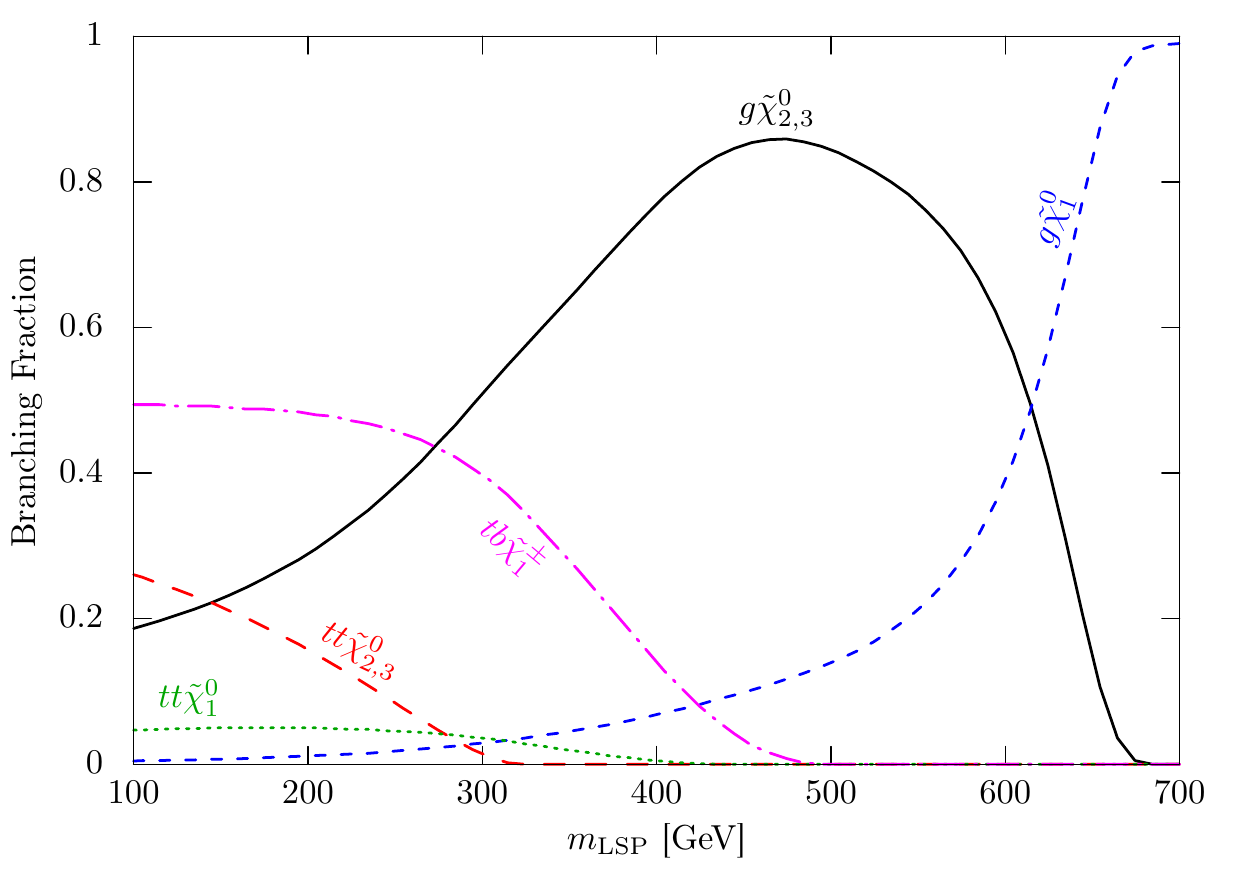}}
 \subcaptionbox{LSP mass dependence of $\text{BF}(\tilde g \to g \tilde\chi^0_{2,3})$\label{subfig:br2}}{\includegraphics[width=0.45\textwidth]{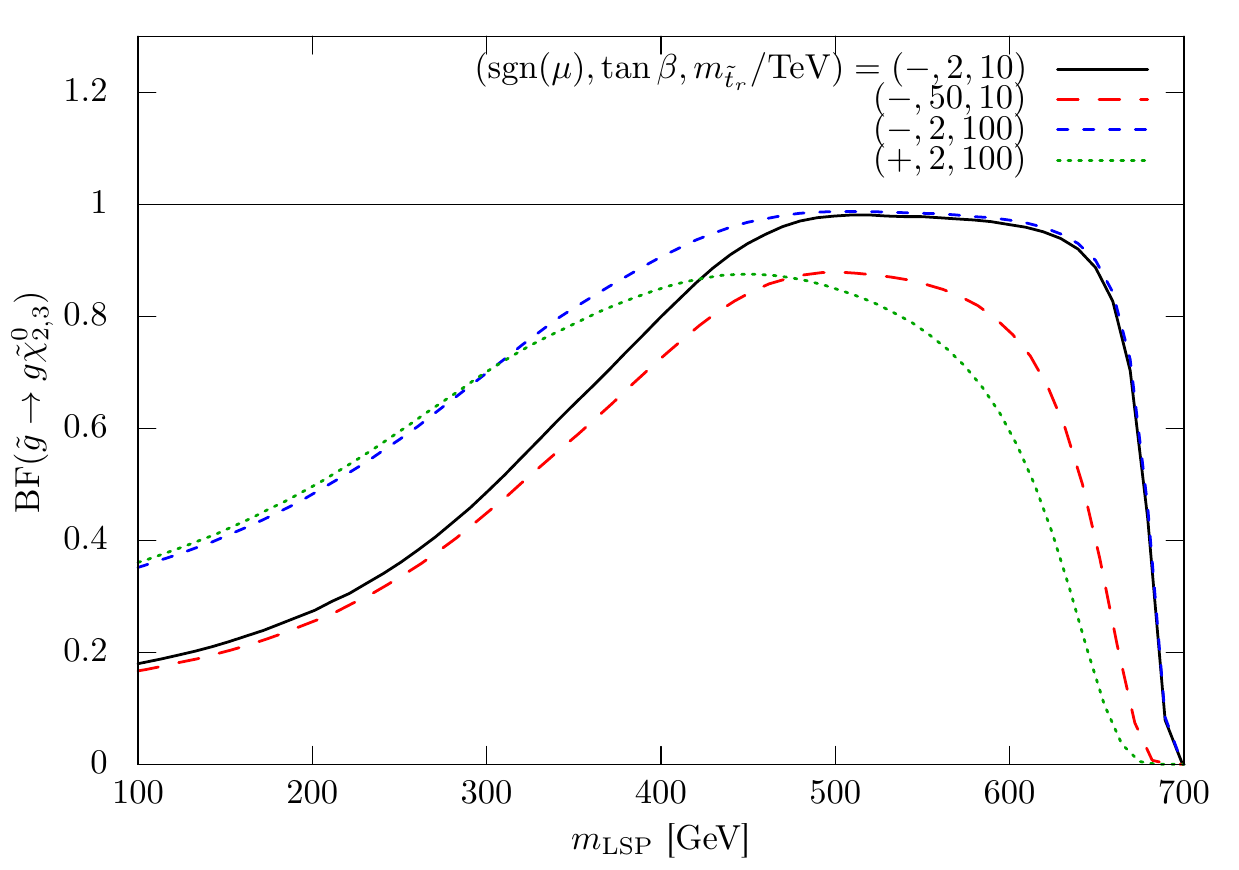}}
 \subcaptionbox{Squark mass dependence of $\text{BF}(\tilde g \to g \tilde\chi^0_{2,3})$ \label{subfig:br3}}{\includegraphics[width=0.45\textwidth]{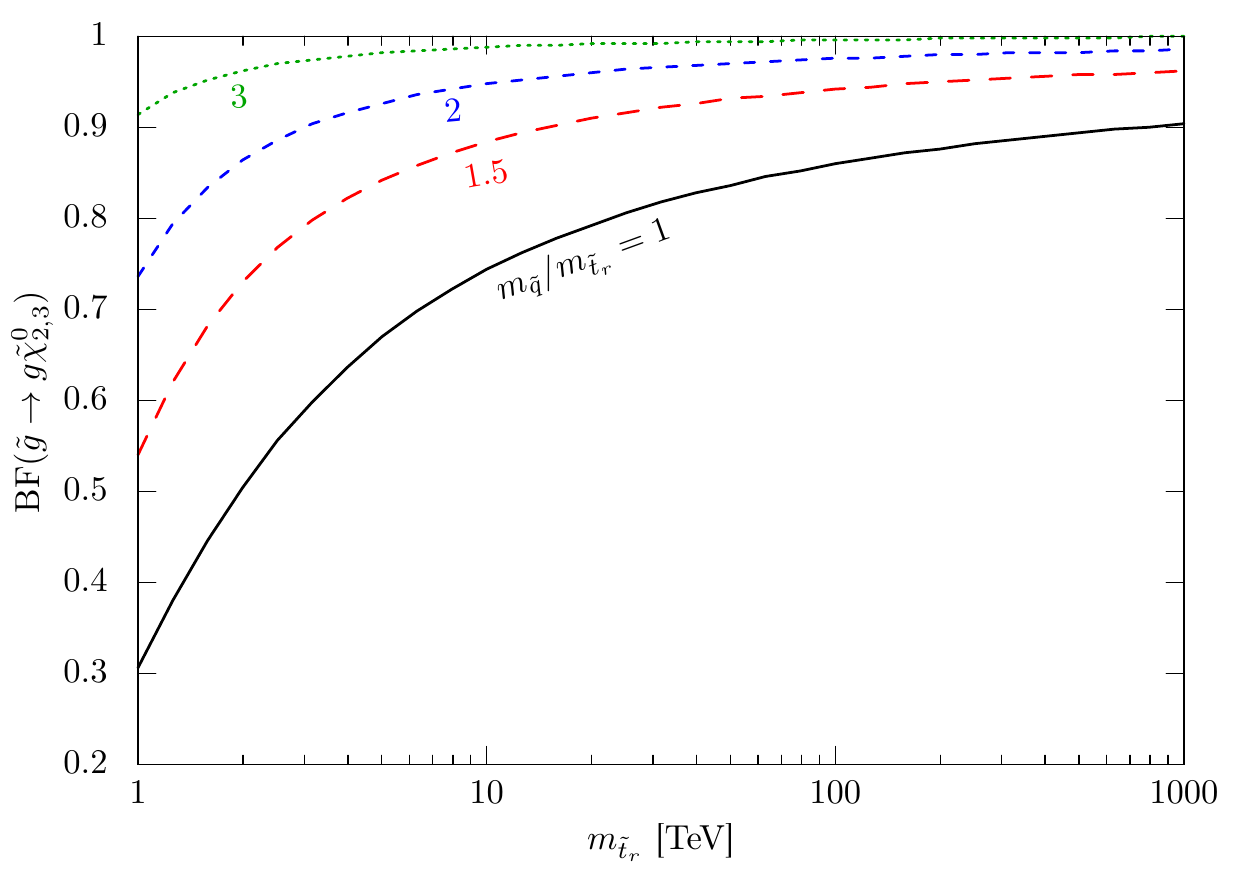}}
 \subcaptionbox{$\text{BF}(\tilde g \to g \tilde\chi^0_{2,3})$ on $m_{\tilde g}$-$m_{\rm LSP}$ plane \label{subfig:br4}}{\includegraphics[width=0.45\textwidth]{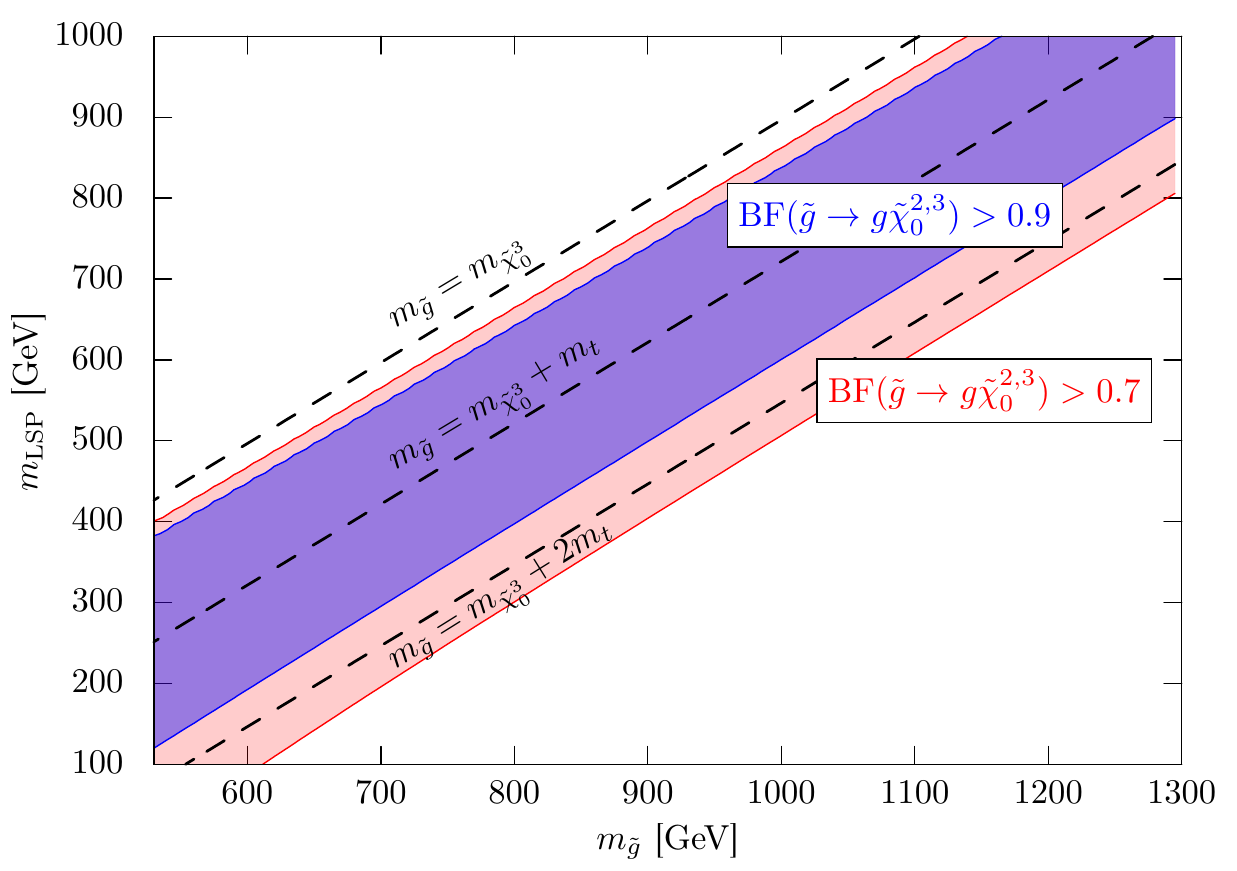}} 
  \caption{Various dependencies of branching fractions of the gluino.
  (a):  The branching fractions of the gluino decay as functions of the LSP mass $m_{\text{LSP}}=m_{\tilde{\chi}_1^0}$. 
The solid black, long-dashed red, medium-dashed blue, short-dashed green, and dot-dashed pink lines show the branching fractions to $g \tilde{\chi}_{2, 3}^0$, $tt\tilde{\chi}_{2, 3}^0$, $g \tilde{\chi}_1^0$, $tt \tilde{\chi}_1^0$, and $tb\tilde{\chi}_1^{\pm}$, respectively.  We take $m_{\tilde{g}}=800$ GeV, and $(\text{sgn}(\mu), \tan \beta, m_{\tilde{t}_r}/\text{TeV})=(+, 2, 10)$.
  (b): The branching fraction into $g \tilde{\chi}_{2, 3}^0$ for various choices of parameters.
  The solid black, long-dashed red, medium-dashed blue, and short-dashed green lines correspond to  $(\text{sgn}(\mu), \tan \beta, m_{\tilde{t}_r}/\text{TeV})=(-, 2, 10), (-, 50, 10), (-, 2, 100),$ and $(+, 2, 100)$, respectively.
  (c): The branching fraction into $g \tilde{\chi}_{2, 3}^0$ as a function of the stop mass $m_{\tilde t_r}$.
  The solid black, long-dashed red, short-dashed blue, and dotted green lines correspond to squarks masses 1, 1.5, 2, and 3 times larger than the lighter stop mass. We set $M_1=500$ GeV, $\mu=-600$ GeV, and tan$\beta=2$.
  (d): The branching fraction of $\tilde g \to g \tilde\chi^0_{2,3}$ on the $m_{\tilde g} - m_{\rm LSP}$ plane.
We set $\tan\beta=2$, ${\rm sgn}(\mu)=-$ and $m_{\tilde t_r}=100$ TeV.  It is greater than 0.7 and 0.9 in the red and blue regions, respectively.
  }
 \label{fig:br}
  \end{figure}

In the calculation of gluino branching fractions, the relevant parameters are gluino mass $m_{\tilde{g}}$, Bino and Wino mass parameters $M_1$ and $M_2$, squark masses, Higgs $\mu$ parameter, CP odd Higgs mass $m_{A}$,  and tan$\beta$.
For simplicity, we neglect mixing among squarks, and we take a universal mass for squarks except for the right-handed stop.  The former is set three times larger than the latter, and $m_{\tilde{g}}=800$~GeV is taken unless otherwise stated.
To reduce the number of parameters, we fix $M_{2}=3$~TeV, and $m_{A}$ same as squark masses. 
To make mass difference between $\tilde{\chi}_1^0$ and $\tilde{\chi}_{2, 3}^0$, $\tilde{\chi}_1^{\pm}$ about 100 GeV, we fix $|\mu|=M_1 + 100$ GeV with $M_1$ positive.
We vary $M_1$ and diagonalize the neutralino (and chargino) mass matrix to show the $m_{\text{LSP}}$ axis in the Figure~\ref{fig:br}, where the branching fractions of gluino are plotted.


In Fig.~\ref{subfig:br1}, five dominant branching fractions of gluino are shown with the right-handed stop mass $m_{\tilde{t}_{r}}=10$~TeV.  The black line represents the branching fraction of the gluino to the Higgsino-like neutralinos and gluon.  It diminishes in the right side of the Figure for the kinematical reason.
As the mass splittings between the relevant neutralino or chargino and gluino increase (to the left of the Figure), the three-body decay channels emitting a quark-antiquark pair become non-negligible.
The choice of the parameters in Fig.~\ref{subfig:br1} is relatively inefficient for the gluon channel.
The branching fraction of the gluino into $g\tilde{\chi}_{2, 3}^0$ for other choices of parameters are shown in Fig.~\ref{subfig:br2}.
The branching fraction becomes large in particular when we take the relative sign between $\mu$ and $M_1$ negative. 
This is because the up-type Higgsino component in the LSP decreases by partial cancellation in the case of low $\tan\beta$ and $\text{sgn}(\mu/M_1)=-1$.

In Fig.~\ref{subfig:br3}, we show the dependence of the gluino branching fraction of $\tilde g \to g \tilde \chi^0_{2,3}$  on the squark masses.
Here we assume $ m_{\tilde q_{L,R 1,2}} = m_{\tilde q_{L3}} \equiv m_{\tilde q}$, $M_{1}=500$ GeV, $\mu = -600$ GeV and $\tan\beta=2$.
We show the cases that $m_{\tilde q}/m_{\tilde t_r} = 1,1.5,2$ and $3$.
If the $m_{\tilde q}/m_{\tilde t_r}$ is large enough, $m_{\tilde t_r} = O(1)$ TeV can realize the favoured gluino decay, which may leave the possibility that the relatively ``natural" SUSY can account for the ATLAS $Z$ excess.
As seen in Fig.~\ref{subfig:br3}, heavier stop can increase the $\text{BF}(\tilde g \to g \tilde\chi^0_{2,3})$, due to the large log enhancement.
Note that, however, the gluino decay length gets larger as the stop mass increases:
\begin{align}
c\tau_{\tilde g} \sim O(10) \mu {\rm m} \left( \frac{m_{\tilde g} - m_{\rm NLSP}}{300~{\rm GeV}} \right)^{-3}  \left( \frac{m_{\tilde t_r}}{100~{\rm TeV}} \right)^{4}.
\end{align}
The standard tracking system assumes the gluino decay occurs within $O(1)$ mm from the primary vertex.
Therefore the stop mass should be less than $O(100)$ TeV to produce the standard MET and/or jets and/or leptons signals.
If the decay length is larger than $O(1)$ mm, severer constraints will be imposed even in the case of the compressed mass spectrum \cite{Nagata:2015hha}.

In Fig.~\ref{subfig:br4}, we also vary the gluino mass as well as the LSP mass, showing the region where the branching fraction of the gluino into a Higgsino-like neutralino and a gluon is high.
Here we set $(\mu/\text{GeV}, \tan \beta, m_{\tilde{t}_r}/\text{TeV})=(-M_1 -100, 2, 100)$, with $M_1$ being real and positive.
We see, $m_{\tilde{g}}-m_{\tilde{\chi}_{2, 3}^0}\lesssim 300$ GeV and $m_{\tilde{\chi}_{2, 3}^0}-m_{\tilde{\chi}_1^0}\simeq 100$ GeV, can lead to efficient $Z$ production, $\text{BF}(\tilde{g}\to g Z \tilde{\chi}_1^0)\simeq 1$.

\section{LHC Signals}  \label{sec:signals}

\begin{figure}[h]
\centering
\includegraphics[width=0.7\textwidth]{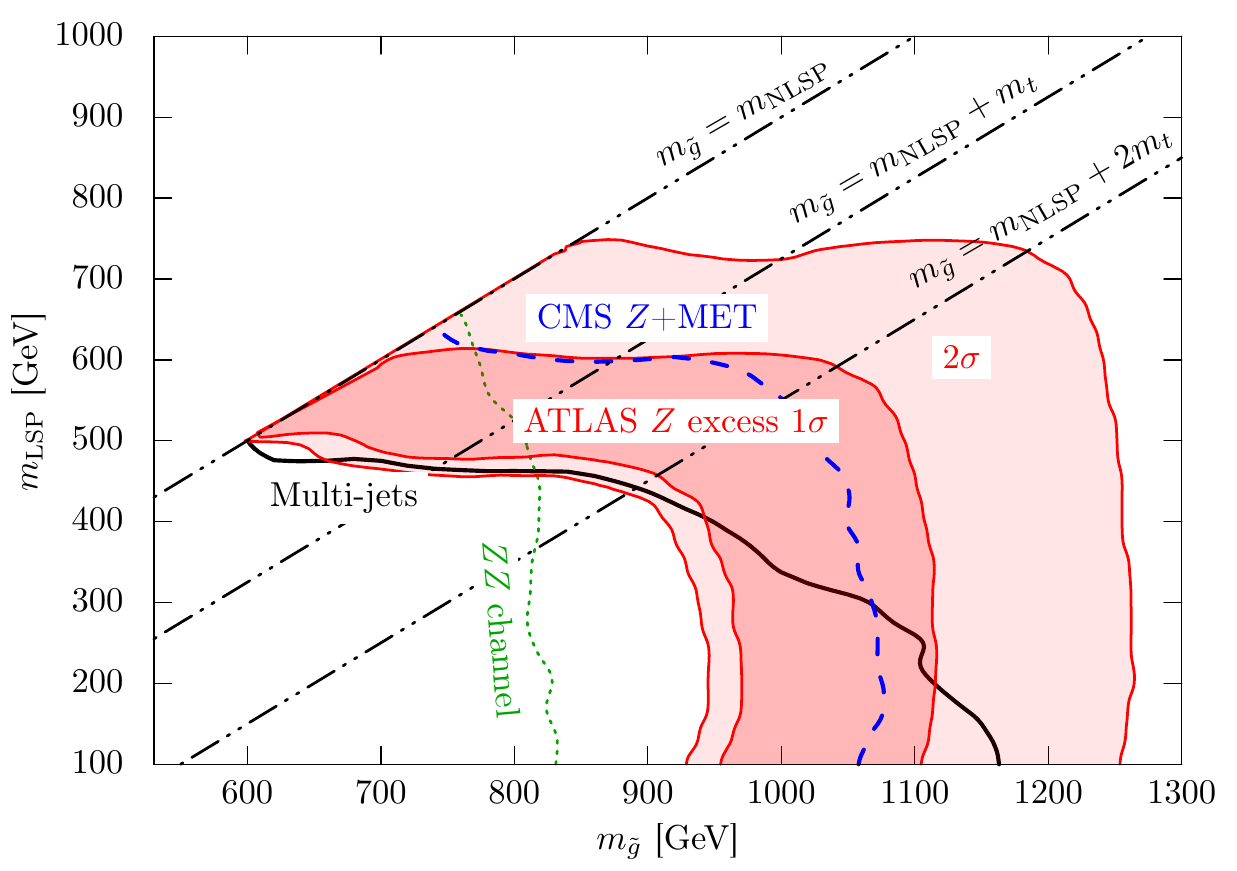}
\caption{
\label{fig:LHC}
Constraints at the $95\%$ CL of the simplified model.
The black solid line shows the ATLAS multi-jets+MET constraints; the blue  dashed line shows the CMS $Z$+MET constraint; and the green dotted line shows a combined $ZZ$ constraint including the ATLAS four lepton, CMS four lepton, and CMS $Z$ $+$ dijet channels.
The red regions show 1$\sigma$ and 2$\sigma$ parameter estimation from the ATLAS excess of the $Z$ channel.}
\label{fig:fit}
\end{figure}


The essential features of the present mass spectrum discussed in the last Section can be reduced into a simplified model with the decay chain: $\tilde g \to g \tilde{\chi}_2^0$ and $\tilde{\chi}^0_2 \to Z \tilde{\chi}^0_1$. We assume $m_{\tilde{\chi}^0_2} - m_{\tilde{\chi}^0_1}=100$ GeV and take the branching fraction of each decay as $1$ for simplicity.

As discussed before, this simplified model can produce sufficient amount of $Z$ bosons, which we expect to be consistent with the ATLAS $Z$ excess. On the other hand, the counterpart search by the CMS collaboration~\cite{Khachatryan:2015lwa} should place severe constraints on it. In addition, this simplified model can easily produce up to six jets due to the gluons produced by the gluino decay and the hadronic decay modes of $Z$ bosons. Therefore, constraints by multi-jets+MET channels could potentially be important. Dilepton+Dijet+MET and four-lepton+MET analyses could also be relevant due to the (semi-)leptonic decay modes of $Z$ bosons. Single lepton+MET channels are expected to be less important due to the second lepton veto.

To study the fitting region and various constraints, we generate this simplified model with up to one extra parton in the matrix element using MADGRAPH 5 v2.1.2~\cite{Alwall:2014hca,*Alwall:2011uj} interfaced to Pythia 6.4~\cite{Sjostrand:2006za} and Delphes 3~\cite{deFavereau:2013fsa} (which has FastJet incorporated~\cite{Cacciari:2011ma,*Cacciari:2005hq}). The MLM matching~\cite{Alwall:2007fs} is applied with a scale parameter set to a quarter of the gluino mass. The parton distribution functions (PDFs) from CTEQ6L1~\cite{Pumplin:2002vw} are used. The gluino production cross sections are calculated at next-to-leading order (NLO) in the strong coupling constant, adding the resummation of soft gluon
emission at next-to-leading-logarithmic accuracy (NLO+NLL) by using NLL-fast v2.1~\cite{Beenakker:1996ch,*Kulesza:2008jb,*Kulesza:2009kq,*Beenakker:2009ha,*Beenakker:2011fu}.

In Fig.~\ref{fig:LHC}, we show $1\sigma$ and $2\sigma$ parameter estimations  from the ATLAS $Z$ excess data~\cite{Aad:2015wqa}. 
For this fitting, we estimate the number of SUSY signal events for the ATLAS cut, which requires a same-flavor opposite-sign dilepton pair with its invariant mass in the $Z$ mass range ($81 \text{\, GeV}< m_{ll} < 101\, \text{GeV}$), two jets, and MET larger than 225 GeV.  Additional cuts include the large scalar sum ($H_{\text{T}}>600$ GeV) of the transverse momenta of all signal jets and the two leading leptons, and large azimuthal angular separations ($\Delta \phi (\text{jet}{}_{1,2}, E_{\text{T}}^{\text{miss}})>0.4$) between each of the leading two jets and the MET direction.
With the observed number 16 (13) and the SM expectation value  $4.2\pm1.6$ 
($6.4\pm 2.2$) for the dielectron (dimuon) channel,  we construct a  $\chi^2$ variable.
The regions $\Delta \chi^2 < $2.3 and 6.0, corresponding to 68th and 95th percentile of the Chi-Squared distribution with
two degrees of freedom, respectively, are referred to as  $1\sigma$ and $2\sigma$ fitting regions in the Figure.

Similarly, we also estimate the signal strength of the  CMS $Z$+MET data~\cite{Khachatryan:2015lwa}, ATLAS multi-jets (2-6 jets)+MET data~\cite{Aad:2014wea} and four-lepton+MET data by ATLAS \cite{Aad:2014iza} and CMS \cite{Chatrchyan:2014aea} and  CMS $Z$+dijet+MET~\cite{Khachatryan:2014qwa}.
Then we estimate the exclusion curves at the $95\%$ confidence level, by using the $CL_s$ prescription.
We use the SM background estimations and its uncertainties provided by each reference.
Regarding the ATLAS multi-jets (2-6 jets)+MET searches, we choose a channel which is expected to give the most stringent constraints on each parameter point among 15 signal regions.
We combine  the  four-lepton+MET data by ATLAS \cite{Aad:2014iza} and CMS \cite{Chatrchyan:2014aea} and  CMS $Z$+dijet+MET~\cite{Khachatryan:2014qwa} and show $ZZ$ channel exclusion line in Fig.~\ref{fig:LHC}.
Other constraints such as ATLAS large jet multiplicities (7-10 jets)+MET~\cite{Aad:2013wta} and ATLAS $Z$+dijet+MET~\cite{Aad:2014vma} are found less important.

In Fig.~\ref{fig:LHC}, the relatively small mass splitting region ($m_{\tilde g}-m_{\text{NLSP}}\lesssim 2m_t$) is of our true interest, because in this region the gluino decay branching fraction in the split MSSM can be very close to $1$ (see Fig.~\ref{fig:br}), and hence justifies our use of this simplified model. We see from the Figure that within the justified parameter region, there is a substantial parameter space that is consistent with ATLAS $Z$ excess and not excluded by the various constraints, apart from the CMS $Z$+MET counterpart search. There is even a small parameter region consistent with both the ATLAS $Z$ excess and the CMS counterpart exclusion limit.

\section{Summary and Discussion} \label{sec:summary}
In this paper, we study the possibility of explaining the recent ATLAS $Z$ excess in the MSSM spectrum.
We study the gluino and neutralino decays in the case that sfermions are heavier than the gluino, assuming the Bino LSP and the Higgsino NLSP.
We show that the small mass difference between the gluino and neutralinos and/or large stop mass can relatively enhance the radiative gluino decay $\tilde g \to g \tilde H_u^0$.
In this model, while the LSP is Bino-like neutralino, the direct gluino decay into the LSP is relatively suppressed.
Motivated by this feature, we explored the simplified model to explain the ATLAS $Z$+MET excess.
We found that the gluino mass  around 800-1000 GeV and $m_{\rm NLSP} \gtrsim m_{\tilde g} - 2 m_{t}$ can well explain the ATLAS $Z$ excess without conflicting with other SUSY searches, apart from the CMS $Z$+MET search.
In such a region the gluino radiative decays into $\chi^0_{2,3}$ are dominated, as seen in Fig.~\ref{fig:br}, and the  present simplified model well describes the realistic gluino decay chains.
Therefore we can conclude the very simple MSSM spectrum may explain the ATLAS $Z$+MET excess.

Moreover the present MSSM mass spectrum has another advantage.
As pointed out in  Ref.~\cite{Ellwanger:2015hva}, the compressed mass spectrum 
can well fit the observed distributions of MET $E_{\text{T}}^{\text{miss}}$ and the scalar sum of transverse momenta $H_{\text{T}}$ in the signal region of the ATLAS $Z$+MET search.
This feature is also the case for the present MSSM model.
The mass spectrum of our interest, thus, can account for not only the number of the ATLAS $Z$ excess without conflicting with the major SUSY searches but also the more detailed behaviours of the excess.

However there is a subtlety when we consider the $Z$+MET search by the CMS, which seems to exclude a large portion of the best-fit region for the ATLAS $Z$+MET signals.
Although the ATLAS and CMS searches are similar to each other, the ATLAS search makes relatively more account of the hadronic activity.
This difference leads to slightly different LHC constraints between the ATLAS and CMS searches.
Then there is a tiny region where the ATLAS excess can be explained and the CMS constraint is evaded.
However the CMS exclusion limit and the ATLAS best fit region is very close, and it is hard to conclude the both searches can be consistent within this model, taking into account possible uncertainties of our fast simulations.
A more detailed detector simulation will be needed to estimate more precise constraints.

The LHC Run II will provide more obvious test for this model.
The production cross section of the gluinos at 13 TeV LHC is enhanced by around 10, compared to the 8 TeV run, assuming the gluino mass is around 1 TeV.
If the dilepton excess really comes from the SUSY particles, the number of the SUSY events for the integrated luminosity $\int{\cal L}dt=10$ fb$^{-1}$ and the center-of-mass energy $\sqrt{s}=13$ TeV will be around 100 with the same event selections as the present ATLAS 8 TeV dilepton plus MET  search.
The main SM background comes from $t\bar{t}$ production and its number is estimated to be around 50.
Assuming the systematic uncertainty of the background estimation is 30\% as
in the case of the 8 TeV result, a 5$\sigma$ or more excess will be observed.
Thus, the LHC Run II will provide a very clear test of the present SUSY model.

It will be worth noting that this mass spectrum may provide a good Bino-like dark matter candidate.
The abundance and detection of the dark matter is quite sensitive to the other parameters, such as the Wino mass, CP phases and $\tan\beta$, and its detailed study is beyond the scope of the present paper.

In this study, we assume the split SUSY-like spectrum in which the scalar tops play a dominant role in the gluino decay, and the Bino and Higgsino are the LSP and NLSP, respectively.
Although this assumption is well motivated,
it is interesting to investigate more generic types of MSSM spectra, and it will be done elsewhere.

\section*{Acknowledgments}

The work of T.T. is supported by a Grant-in-Aid for JSPS Fellows, and a JSPS Grant-in-Aid for Scientific Research No.~26$\cdot$10619.

\bibliographystyle{aps}
\bibliography{ref}

\end{document}